# Allometric metabolic scaling and fetal and placental weight

Allometric scaling in fetus and placenta

Carolyn M. Salafia MD MS[1] Dawn P. Misra, PhD[2], Michael Yampolsky, PhD[3]

Adrian K. Charles, MBChB[4], Richard K. Miller, PhD[5]

[1]Department of Obstetrics and Gynecology, St. Luke's Roosevelt Hospital, New York, New York, Carolyn.Salafia@gmail.com.

[2] Department of Epidemiology, University of Michigan, Ann Arbor, MI, dmisra@umich.edu.

[3] Department of Mathematics, University of Toronto, Toronto, CA, Yampolsky.michael@gmail.com.

[4] Department of Pathology, Princess Margaret Memorial Hospital, Perth WA, Adrian.Charles@optusnet.com.au.

[6] Department of Obstetrics and Gynecology, University of Rochester, Rochester, NY, RichardK_Miller@urmc.rochester.edu.

Corresponding author:

Carolyn M. Salafia MD MS

St. Luke's Roosevelt Hospital

Department of Obstetrics and Gynecology

1000 Tenth Avenue

New York, New York 10019

Email: Carolyn.salafia@gmail.com

Number of characters in abstract 887

Number of characters in entire manuscript 27,853

ABSTRACT

Background: We tested the hypothesis that the fetal-placental relationship scales allometrically and identified modifying factors.

Methods: Among women delivering after 34 weeks but prior to 43 weeks gestation, 24,601 participants in the Collaborative Perinatal Project (CPP) had complete data for placental gross proportion measures, specifically, disk shape, larger and smaller disk diameters and thickness, and umbilical cord length. The allometric metabolic equation was solved for α and β by rewriting PW= α (BW)^β as Log (PW) = Log α + β [Log(BW)]. α was then the dependent variable in regressions with $p<0.05$ significant.

Results: Mean β was 0.78± 0.02 (range 0.66, 0.89), 104% of that predicted by a supply-limited fractal system (0.75). Gestational age, maternal age, maternal BMI, parity, smoking, socioeconomic status, infant sex, and changes in placental proportions each had independent and significant effects on α.

Conclusions: In the CPP cohort, the placental - birth weight relationship scales to ~a ¾ power.

## INTRODUCTION

The human newborn is entirely composed of nutrients transferred from the maternal to the fetal circulation across the placenta. By extension, birth weight depends on placental function. Physiologic determinants of total placental supply capacity include “driving forces” such as nutrient concentration, charge and oncotic gradients, blood flow (via uterine and umbilical arteries), the physical aspects of the placental villous barrier related to passive permeability (e.g., villous surface area, thickness of the maternal-fetal blood partition, pore size), and transporter function at the microvillous surface. The “net” (or “effective”) placental functional capacity perceived by the fetus would equal the amount of nutrients provided in each fetal-placental cardiac cycle minus the fetal energy costs of placental perfusion and the energy consumed by placental metabolism. [1]

The relationship between metabolism and organism size has been repeatedly reduced to the equation **Basal metabolic rate= $\alpha$(Body mass)^$\tilde{\beta}$** (e.g., [2-9]) Allometric metabolic scaling (Kleiber’s Law) was first conjectured in the 1930’s and has proved to be remarkably constant for a wide range of organisms from the smallest microbes ($\sim 10^{-13}$ g) to the largest vertebrates and plants ($\sim 10^{8}$ g, [9]). Allometric scaling is applied to extrapolate human toxic doses from results of experimental models and is remarkably constant, between 0.67 and 0.75. Such scaling applies in growing as well as mature organisms. [3] Recent models suggest that allometric scaling can be understood on the basis of supply limitations [9] or as the combined effects of processes that each contribute to regulation of whole-body metabolic rate [6-8]. In either case, the broad range of observed data underlying the concept that functional capacities are matched to maximum physiological requirements or loads (“symmorphosis” [10]) has been used to propose that such balance was a basic evolutionary requirement.

Ahern in 1966 (as cited in [11]) offered a pregnancy-equivalent to the allometric metabolic scaling equation that suggested that the complex relationship between placental function in nutrient transfer and fetal growth could also be parsed allometrically. He substituted placental and fetal masses for basal metabolic rate and body mass, respectively, yielding the following equation:

$$\text{Placental weight} = \alpha(\text{Birth weight})^{\wedge}\beta$$

He suggested that, based on a “series of ‘normal’ products of conception” that β=2/3, consistent with scaling based on volumes and surface areas. However, recently a ¾ scaling has been favored, based on flow theory, in which the rate of delivered materials to cells by the circulation, is constrained by the capacity to deliver materials, or fractal theory (that justifies a ¾ power as the effective surface area optimal to achieve maximum metabolic rate while minimizing internal transport distance) [2-4]. Placental nutrient transfer to the fetus is intimately dependent on placental flow; placental branched growth is essentially fractal. [12] Thus either flow or fractal theory might be applicable to the fetal-placental version of allometric scaling, both supporting a ¾ scale factor.

As investigations of the putative “fetal origins of adult disease” have proliferated, so has the use of birth weight as a proxy for intrauterine “health” (as recently summarized in [13]). Birth weight is currently understood to be a marker of risk for a host of lifelong health risks, but evidence suggests that birth weight per se is not the critical exposure. [14] “Fetal origins” investigations would benefit from a clearer understanding of how the placenta, a principal determinant of fetal growth, “translates” its own growth into fetal mass. While reduced placental growth is generally associated with reduced fetal growth, structural changes in the placental vascular distribution system (e.g., abnormal coiling [15], single umbilical artery[16], abnormal umbilical cord insertion[17]) are associated with increased rates of fetal growth restriction. It is likely that finer-level variations in the anatomic architecture of the placental tree, from the larger branches of the umbilical chorionic and fetal stem vessels as well as the distal capillary bed, would also affect the fetal "net benefit" of placental perfusion and alter the birth weight for a given placental weight.

The finer structure of the placenta is difficult to quantify but several proxies exist in a large US birth cohort, the Collaborative Perinatal Project (CPP) [18]. First, measures of the larger and smaller dimension of the chorionic disk describe the area of the chorionic plate which would constrain the maximal length of the chorionic plate vessels (and potentially affect fetal cardiovascular work since cardiovascular resistance is directly proportional to vessel length). Second, measures of placental disk thickness offer a crude quantification of the depth/extent of villous arborization, not only in terms of villous nutrient exchange surface area but also the fetal stem arterioles, the principal site of placental vascular resistance. Altered proportions of the chorionic plate (a very small chorionic plate area or a very large and expansive chorionic plate) or placental disk thickness (a thin or thick disk) would therefore, in theory, represent very different fetal-placental relationships. The more fetal work involved in perfusing the placenta, the greater "cost" to the fetus of every heart beat and the lesser would be, in theory, the "net" nutrient benefit of each fetal cardiac cycle.

We first hypothesize that the mean allometric exponent relating placental and fetal size will approximate that predicted by both flow and fractal theories, ¾. If this is true, we then hypothesize that changes in placental three-dimensional shape, the container of the mature and arborized villous tree (estimated by measures of the smaller and larger placental diameters as proxies for the placental chorionic vascular distribution system, and placental disk thickness as proxy for the elaboration of the villous distribution system) will alter the balance between placental weight and fetal weight, in other words, affect the efficiency of placental nutrient transport function. Through the allometric metabolic scaling equation, we can test this hypothesis by assessing whether, and if so, to what degree, these placental growth variables affect the linear coefficient α of the allometric metabolic scaling equation. Finally, we reconsider what is the most biologically appropriate expression of the feto-placental weight ratio, in light of scaling considerations. The implications of scaling are significant, as a linear fetoplacental weight ratio (birth weight/placental weight) is a commonly used clinical tool for assessing fetal well-being and placental health. If a corrected ratio should be used (e.g., birth weight $^{3/4/}$placental weight), this would refine clinical diagnosis of fetal pathology due to placental dysfunction. Deviations from the ratio ¾ could thus also be used as indicators of deviations from normal fetal-placental development.

METHODS:

Subjects were a subset of the National Collaborative Perinatal Project (NCPP). Details of the study have been described elsewhere [19, 20]. Briefly, from 1959 to 1965, women who attended prenatal care at 12 hospitals were invited to participate in the observational, prospective study. At entry, detailed demographic, socioeconomic and behavioral information was collected by in-person interview. A medical history, physical examination and blood sample were also obtained. In the following prenatal visits, women were repeatedly interviewed and physical findings were recorded. During labor and delivery, placental gross morphology was examined and samples were collected for histologic examination. The children were followed up to seven years of age.

The data for the present analysis was derived from all liveborn singletons. To control for correlated observations the sample was restricted to only or first singleton live births within a family. Among 41,970 women who provided eligible singleton births, 36,017 contributed placenta data. The sample was further restricted to those with complete data on the six placental gross measures (described below), placental weight, and birth weight, of known gestational age between 34 weeks and 42 6/7 (less than 43) completed weeks (N=24,152). Gestational age was calculated from the last menstrual period in rounded weeks. The 34 week cut off was selected because gestations younger than this age were less likely to survive. Gestational lengths as great as 54 weeks were also reported; these were clearly in error and were also excluded. Finally, the lowest 0.5% of placental variables was excluded, as biologically implausible, leaving a final analytic sample of 24,061. There were no exclusions for diagnoses of diabetes, preeclampsia or other maternal medical conditions.

Placental gross measures included placental disk shape, relative centrality of the umbilical cord insertion, estimated chorionic plate area, disk eccentricity, placental disk thickness, placental weight, and umbilical cord length, measured according to a standard protocol [18]. Gestational age was calculated based on the last menstrual period in rounded weeks. Among 41,970 women who gave the first or only singleton live birth, 36,017 contributed placenta data. The analytic sample was restricted to those with complete data on the six placental gross measures, placental weight and birth weight, of gestational ages >= 34 weeks (younger infants having been unlikely to survive) and less than 43 completed weeks (given that gestations were assigned implausible gestational lengths up to 54 weeks, N=24,061). The original coding of placental measures and the recoding used for this analysis follow:

- Chorionic disk shape coding was based on the gross examination of the delivered placenta. Shapes included round-to-oval, and a variety of atypical shapes (e.g., bipartite, tripartite, succenturiate, membranous, crescent or "irregular"). Only 926 (3.8 percent) were labeled as one of the 6 categories of shape other than round-to-oval. For this analysis, the shape measure was recoded as a binary variable with "round-to-oval" as "0" and "other than round-to-oval" as "1".
- Relative centrality of the umbilical cord insertion was calculated from two variables recorded in the original data set. The distance from the cord insertion to the closest placental margin was recorded to the nearest cm. The type of umbilical cord insertion was coded as membranous (velamentous), marginal or normal (inserted onto the chorionic disk). We combined these two variables into a single distance measure, by recoding velamentous cord insertions as a negative value, cords inserted at the placental margin as "0" and progressively more central cords as "1" to "9" (overall scale range -13 to 13).

- Estimated chorionic plate area was calculated as the area of an ellipse from two variables recorded in the original data set, the larger diameter and smaller diameter of the chorionic disc were recorded in cm. Disk eccentricity was calculated as the ratio of the larger and smaller diameters. Both the chorionic plate area and disk eccentricity could be cast as "interactions" between larger and smaller disk diameters.
- Placental thickness at the center of the chorionic disc was recorded in units of 0.1 cm, by piercing the disc with a knitting needle on which millimeter marks were inscribed.
- Placental weight was measured in decagrams to the nearest 10 grams; this variable was converted to grams.
- The fetoplacental weight ratio was calculated as birth weight divided by the placental weight, and is a value generally considered to reflect a physiologic state of balance between fetal and placental growth.
- Umbilical cord length was analyzed as it was measured in the Labor and Delivery Room. Cord lengths ranged from seven to 98 cm.

Maternal characteristics were recorded at enrollment. Maternal age was coded as age at (enrollment) in years, and maternal height was measured in inches. Maternal weight prior to pregnancy was self-reported in pounds. Body mass index (BMI) was calculated from maternal height and weight. Parity counted all delivered live born offspring and did not include miscarriages/early pregnancy losses. Socioeconomic status index was a combined score for education, occupation and family income as scaled by the US Bureau of the Census. [21] Mother's race was coded as a binary variable denoting African-American as "1" and all others as "0"; original data coded race as Caucasian, African American, and "other", most of whom were Puerto Ricans (9.2 percent). Cigarette use was coded by maternal self report at enrollment as non-smoker (coded as <1 cigarette per day), or by the self-reported number of cigarettes smoked daily grouped as 1-9, 10-20, and >20 (greater than one pack per day).

Solving the metabolic scaling equation:

We first solved the allometric metabolic equation for estimates of $\alpha$ and $\beta$. Specifically, PW= $\alpha(BW)^\beta$ is rewritten as a standard regression equation and solved for $\alpha$ and $\beta$:

$$\text{Log (PW)} = \text{Log } \alpha + \beta \text{ Log (BW)]} \quad \text{[Equation 1.1]}$$

From Equation 1.1,

$$\text{Log } \alpha = \text{Log (PW)} - \beta \text{ Log (BW)]} \quad \text{[Equation 1.2];}$$

Substituting the mean $\beta$ for the population, this second equation was solved for each case, and the calculated Log $\alpha$. $\alpha$ was exponentiated and used as a dependent variable in subsequent analyses.

Testing for significant influences on $\alpha$

Spearman's rank correlations and multivariate regression were used to determine significant associations with $\alpha$. $P<0.05$ was considered significant throughout. Three analyses were run. The first included all placental variables; thus the point-estimate of effect for each placental variable is adjusted for the presence of the others. The second included all maternal and fetal variables; again, data presented reflect effects adjusted for the presence of the other maternal variables. The third analysis included all variables (placental, maternal and fetal).

## RESULTS

### Descriptive Statistics

Population descriptors are presented in Table 1 and Table 2.

### Solving for α and β:

The mean (exponentiated) α and β were 1.03± 0.17 (range 0.38, 2.42), and 0.78± 0.02 (range 0.66, 0.89) the latter 104% of the allometric exponent predicted in a supply-limited fractal system, 0.75. [2, 4]

### Placental gross growth dimension variables: effects on α (Table 3, Column 1)

After adjustment for placental weight, α was less in irregularly shaped placentas; by contrast, α increased as placental larger and smaller placental diameters, disk thickness and umbilical cord length increased. Chorionic plate area, a form of interaction between larger and smaller diameters, was inversely correlated with α, suggesting a negative effect on placental efficiency at extreme chorionic plate sizes. Thus, the placental dimensions of shape, chorionic plate (area and the larger and smaller diameters individually), disk thickness, and cord length had separate significant effects on the extent to which placental weight represents fetal metabolic rate. The placental gross growth dimensions accounted for 24% of α variance (r=0.49).

### Maternal and fetal factor effects on α (Table 3, Column 2)

As gestational age and maternal BMI increased, α decreased. Higher doses of maternal smoking was associated with greater α. Male infants had lower α than female infants. No difference in α was seen between Caucasians and African-Americans. α was positively correlated with birth length and inversely correlated with maternal BMI (r=0.035 and r=0.08, respectively). Gestational age also was inversely correlated with α. All the maternal and fetal factors considered in this analysis accounted for 3.6% of α variance (r=0.19).

Do placental measures mediate effects of maternal and infant variables on $\alpha$ (Table 3, Column 3)?

After adjustment for placental variables, increasing maternal age, parity and African-American race (not associated with $\alpha$, absent inclusion of placental variables, Column 1) were significantly associated with $\alpha$, indicating that these variables impact $\alpha$ via effects on one or more placental variables. In particular, the point-estimate of effect of African-American race increased 10-fold after inclusion of placental variables (0.002, p ns, v. 0.023, $p<0.0001$). The effect of birth length on $\alpha$ reversed sign and increased 5-fold (0.020, $p<0.0001$ v. -0.011, $p<0.0001$). The effect of maternal smoking on $\alpha$ was little changed by inclusion of placental variables (0.022, $p<0.0001$ v. 0.017, $p<0.0001$).

What is the most appropriate mathematical expression of the birth weight- placental weight relationship?

In earlier published work examining effects of chorionic disk thickness and area, we reported a difference of almost 30% between the fetoplacental weight ratios observed in the smallest and thinnest placentas and the largest and thickest placentas (8.1 and 6.0, respectively,[22]). Recalculating the fetoplacental weight ratio as birth weight/ placental weight ^.75, as suggested by the results described in this manuscript, essentially eliminates the variability in fetoplacental weight relationships in differently proportioned placentas. (Table 4)

DISCUSSION:

The allometric exponent, and the relationship of placental weight to birth weight:

Our results demonstrate that in this population predominantly delivered at term, placental weight scales to birth weight to the ¾ power, from which we suggest the following:

1. Placental weight is a justifiable proxy for fetal metabolic rate when other measures of fetal metabolic rate are not available.
2. The allometric relationship between placental and birth weight implies that the fetal-placental unit functions as a fractal supply limited system.

To our knowledge, only one other study of placental allometry has been carried out [23] that focused on the allometry of various placental compartments from 10-41 weeks gestation. This work supports our findings that there exist various fetal-placental allometric scaling relations different from a simple proportionality.

We have recently developed a dynamical model of placental vascular growth [12], which relates the spatial shape of the placenta with the structure of the underlying vascular fractal. This model recapitulates many of the known variant placental shapes (bilobate, multilobate, irregular or "scalloped") by a change of the fractal structure of the vasculature at a specific time instance. In a modern cohort with more precise chorionic plate measures than were collected in the CPP, we have confirmed that such placental shape variance is associated with changes in the allometric scaling relation of fetus and placenta. Specifically, deviations from spatially filling symmetric fractal vascular growth are associated with reduced placental vascular efficiency, that is, a smaller birth weight than predicted by the allometric scaling for the given placental weight. These findings confirm the connection between the allometric fetoplacental scaling and the fractal structure of the vascular supply system.

We anticipate that future investigations may further improve these estimates of placental function by including other measures of placental growth, e.g., arborization density, enzyme activity (17 β hydroxysteroid dihydrogenase), and microvillus transport capacity.) We explicitly included all CPP cases with complete placental data; we did not exclude cases of diabetes, preeclampsia or abruption, as one purpose of the present effort was to determine whether, as a general principle, placental weight scaled to birth weight. Deviations from a ¾ scale in the context of maternal medical conditions may help better understand how maternal diseases, via effects on placental growth and, by extension, placental function affect the fetus.

Optimal transport and placental anatomy:

Placental arborization is an essentially random process of fractal growth influenced by permissive and restrictive genetic and environmental factors. Our results suggest that, in general terms, the relationship between placental structure and placental function (in terms of nutrient transport allowing fetal growth) fits an allometric scaling model. The apparent universality of scaling among living organisms has been tied to the idea that evolution also drives optimality of structure, such an "optimal structure" having no excess structures relative to its maximal function (e.g., $O_2$ flux in the lungs, blood flow in the vascular tree, etc.)[10] Optimization theory has been used to analyze a number of biological relationships over the years, from feeding strategies to locomotor gaits (recently reviewed by [24]). In this light, our findings suggest that, in general, and in a predominantly term birth cohort, placental structure is optimized. Likewise, in this light, our findings that changes in placental shape and placental dimensions—independent of their associations with placental weight—affect the balance between placental weight and birth weight can be interpreted as those changes in shape and dimension reflecting deviations from optimal placental structure.

We speculate that these changes in placental shape and dimension are the physical manifestations of altered placental growth necessitated by the intrauterine environment. Using more precise measures of the chorionic shape, we have demonstrated that the radial standard deviation from the umbilical cord insertion is significantly correlated with β [25], further albeit indirect evidence that abnormal chorionic surface perimeters, the "errors in outline" acknowledged by Drs Benirschke and Kaufmann, [26] reflect a placental architecture in which function (of nutrient and oxygen transport) is no longer maximized.

Influences on placental-fetal scaling:

Placental gross growth measures and several maternal characteristics influence placental-fetal scaling. An increase in α implies a larger placenta for a given birth weight, and a lower fetoplacental weight ratio, and a smaller birth weight for any given placental weight. The "optimal" result of placental growth should be to yield greater fetal nutrient transfer and a larger baby, rather than a larger placenta. Our data suggest that the maternal and fetal variables we examined have at least part of their effects on the normal balance between placental weight and birth weight via effects on gross placental growth dimensions. As noted above, the fact that placental growth parameters also affect α independent of placental weight is consistent with our hypothesis that early gestational constraints that yield variant shapes and dimensions of the mature arborized placental villous tree have a permanent effect on the delivered birth weight. Gestational age also showed a significant effect on scaling, despite the admittedly problematic estimation of gestational age by last menstrual period in this cohort. Gestational age should not, in and of itself , affect the ¾ power relationship between placental weight and birth weight. That there is a significant association of even poorly measured gestational age on the scaling relationship suggests that the pathology(ies) that underlie shortened gestational lengths may have (chronic) effects on placental vascular/fractal structure. This is consistent with many studies that have associated chronic placental pathologies with preterm birth. [27-31]

Should the fetoplacental weight ratio calculation be modified to scale placental weight to the ¾ power?

The ratio of birth to placental weight is the common yardstick used in clinical assessment of the appropriateness of placental function (in terms of providing the fetus with nutrients and allowing fetal growth) to the placental mass. Our data suggest that the relationship between birthweight as a measure of placental function and placental weight is not linear but instead scales to the ¾ power predicted by both flow or fractal theories. While this calculation is more cumbersome, deviations from ranges presented in Table 4 may be most clinical precise as they would identify placentas with truly altered flow patterns or fractal structure.

In summary, data from the Collaborative Perinatal Project demonstrate that placental weight and birth weight in the mid-late third trimester scale consistent with allometric scaling power laws. We hypothesize that maternal and/or fetal pathologies (e.g., preeclampsia) known to modify either branch calibers or the branching structure per se will yield "suboptimal" placentas, in terms of birth weight. Better characterization of the branching growth of the placenta may be facilitated by allometric modeling, to develop computer models of placental structure that reflect placental function, and potentially to provide a feasible model for branching growth of other fetal viscera.

ACKNOWLEDGEMENT

This study was supported by a Young Investigator Award from the National Alliance for Research on Schizophrenia and Depression, Great Neck, NY (Dr Salafia); grants (1 K23 MH067857-01) Mid Career Development Award from the National Institutes of Mental Health (Dr. Salafia).

e

**TABLE 1. Descriptives of the National Collaborative Perinatal Project subsample with placental data, children in placenta analysis (N=24,061).**

| | | Number (%) | Birth Weight Mean (SD) | Placental Weight Mean (SD) | Disk Shape "not round" Number (%) | Larger Diameter (cm) Mean (SD) | Smaller Diameter (cm) Mean (SD) | Disk Thickness (mm) Mean (SD) | Umbilical Cord Length (cms) Mean (SD) | Distance from cord insertion to nearest disk margin (cms) Mean (SD) |
|---|---|---|---|---|---|---|---|---|---|---|
| **Maternal race** | White | 11713 (48.7) | 3285.1 (489.0) | 446.1 (92.4) | 576 (4.9) | 19.2 (2.1) | 16.5 (1.9) | 23.3 (4.1) | 59.3 (13.3) | 4.6 (2.3) |
| | Black | 10134 (42.1) | 3095.6 (483.2) | 426.3 (90.0) | 275 (2.7) | 18.9 (2.2) | 16.4 (1.9) | 20.3 (4.6) | 58.7 (13.3) | 5.1 (2.2) |
| | Other | 2214 (9.2) | 3161.2 (465.3) | 435.6 (91.4) | 79 (3.6) | 19.0 (2.2) | 16.2 (1.9) | 20.9 (4.5) | 57.0 (13.0) | 4.8 (2.2) |
| **Maternal socioeconomic status (SES)** | 0-1.9 | 1660 (6.9) | 3087 (487.0) | 420.8 (90.0) | 45 (2.7) | 18.8 (2.1) | 16.3 (1.9) | 20.5 (4.6) | 57.6 (13.1) | 5.0 (2.2) |
| | 2.0-3.9 | 6968 (29.0) | 3147.4 (498.2) | 432.3 (92.1) | 200 (2.9) | 18.9 (2.2) | 16.4 (1.9) | 20.9 (4.6) | 58.1 (13.3) | 5.0 (2.2) |
| | 4.0-5.9 | 7156 (29.7) | 3183.7 (492.4) | 438.3 (92.7) | 267 (3.7) | 19.1 (2.1) | 16.4 (1.9) | 21.6 (4.6) | 58.9 (13.4) | 4.9 (2.2) |
| | 6.0-7.9 | 4827 (20.1) | 3239.9 (487.5) | 442.6 (91.7) | 211 (4.4) | 19.1 (2.1) | 16.4 (1.9) | 22.7 (4.4) | 59.8 (13.3) | 4.7 (2.3) |
| | 8.0-9.5 | 3048 (12.7) | 3311.3 (454.8) | 442.4 (87.4) | 186 (6.1) | 19.3 (2.0) | 16.6 (1.8) | 24.0 (3.8) | 60.0 (12.9) | 4.5 (2.4) |
| | Unknown | 402 (1.7) | 3181.6 (551.2) | 441.6 (98.8) | 21 (5.2) | 19.3 (2.2) | 16.8 (2.0) | 21.1 (4.6) | 56.6 (12.8) | 5.0 (2.4) |
| **Previous births (parity)** | None | 9515 (39.6) | 3152.1 (469.5) | 427.9 (87.3) | 304 (3.2) | 19.0 (2.0) | 16.3 (1.9) | 22.0 (4.4) | 58.4 (13.1) | 4.7 (2.2) |
| | 1 | 5062 (21.0) | 3208.9 (471.4) | 440 (90.5) | 202 (4.0) | 19.0 (2.1) | 16.5 (1.9) | 21.9 (4.7) | 58.2 (13.0) | 4.9 (2.3) |
| | 2 | 3334 (13.9) | 3200.9 (499.3) | 439.6 (92.5) | 157 (4.7) | 19.1 (2.2) | 16.5 (1.9) | 21.8 (4.7) | 59.5 (13.6) | 4.8 (2.4) |
| | 3-9 | 5995 (24.9) | 3241.5 (534.5) | 446.2 (97.7) | 261 (4.4) | 19.1 (2.3) | 16.5 (1.9) | 21.5 (4.8) | 59.7 (13.6) | 5.0 (2.3) |
| | ≥10 | 127 (0.5) | 3321.4 (553.3) | 459.5 (102.0) | 5 (3.9) | 19.5 (2.4) | 16.6 (2.1) | 20.9 (5.1) | 60.1 (14.0) | 5.0 (2.2) |
| | Unknown | 28 (0.1) | 3073.9 (466.8) | 443.2 (89.5) | 1 (3.6) | 18.6 (2.0) | 16.6 (2.3) | 23.3 (4.2) | 60.5 (15.6) | 5.3 (2.2) |
| **Maternal cigarette use (cigarettes/day)** | Non-smoker | 13132 (54.6) | 3268.5 (483.7) | 436.8 (92.0) | 519 (4.0) | 19.0 (2.1) | 16.4 (1.9) | 21.9 (4.7) | 59.2 (13.5) | 4.8 (2.3) |
| | 1-9 | 4262 (17.7) | 3121.4 (492.2) | 430.2 (88.9) | 145 (3.4) | 19.0 (2.2) | 16.4 (1.9) | 21.3 (4.6) | 58.3 (13.1) | 4.9 (2.2) |
| | 10-19 | 3047 (12.7) | 3101.8 (481.8) | 439.8 (93.7) | 112 (3.7) | 19.1 (2.1) | 16.5 (1.9) | 21.9 (4.5) | 58.5 (13.0) | 4.9 (2.2) |
| | >=20 | 3490 (14.5) | 3080.4 (488.1) | 442.3 (92.3) | 145 (4.2) | 19.2 (2.1) | 16.5 (1.9) | 22.2 (4.4) | 58.5 (13.1) | 4.8 (2.3) |
| | Unknown | 130 (0.5) | 3239.5 (521.5) | 437.9 (88.5) | 9 (6.9) | 19.3 (2.4) | 16.6 (2.1) | 21.9 (4.5) | 56.7 (12.6) | 5.1 (2.0) |
| **Maternal Age** | <20 | 5967 (24.8) | 3113.5 (472.1) | 426.3 (87.9) | 124 (2.1) | 18.8 (2.0) | 16.3 (1.8) | 21.5 (4.5) | 57.6 (13.1) | 4.9 (2.2) |
| | 20-29 | 13840 (57.5) | 3208.6 (481.6) | 437.3 (90.7) | 520 (3.8) | 19.0 (2.1) | 16.4 (1.9) | 22.0 (4.6) | 58.7 (13.1) | 4.8 (2.3) |
| | 30-39 | 3882 (16.1) | 3262.3 (535.6) | 450 (97.0) | 252 (6.5) | 19.5 (2.3) | 16.7 (2.0) | 21.8 (4.8) | 61.1 (13.8) | 4.9 (2.3) |
| | >=40 | 372 (1.6) | 3221.6 (594.6) | 446.9 (113.7) | 34 (9.1) | 19.6 (2.7) | 16.7 (2.3) | 21.6 (5.0) | 61.0 (13.6) | 4.8 (2.5) |
| **Maternal weight** | <100 | 1423 (5.9) | 2945.6 (457.4) | 400.4 (80.1) | 60 (4.2) | 18.7 (2.1) | 16.1 (1.9) | 21.0 (4.4) | 54.8 (12.3) | 4.7 (2.3) |
| | 100-149 | 18214 (75.7) | 3178.6 (477.8) | 433.6 (89.3) | 712 (3.9) | 19.0 (2.1) | 16.4 (1.9) | 21.8 (4.5) | 58.5 (13.1) | 4.8 (2.3) |
| | 150-199 | 3381 (14.1) | 3353.3 (500.5) | 463.3 (96.9) | 112 (3.3) | 19.4 (2.2) | 16.7 (1.9) | 22.1 (4.9) | 62.0 (13.7) | 5.0 (2.3) |
| | >=200 | 483 (2.0) | 3443.3 (573.8) | 480.2 (107.6) | 21 (4.4) | 19.5 (2.2) | 16.8 (2.0) | 22.2 (4.8) | 63.8 (14.2) | 5.1 (2.4) |
| | Unknown | 560 (2.3) | 3144.6 (556.3) | 436.6 (101.4) | 24 (4.3) | 19.0 (2.3) | 16.2 (2.0) | 22.0 (5.2) | 58.5 (13.4) | 4.6 (2.4) |
| **Maternal BMI** | <20 | 6067 (25.2) | 3084.5 (475.2) | 417.6 (85.8) | 226 (3.7) | 18.8 (2.1) | 16.2 (1.9) | 21.6 | 57.1 (13.0) | 4.7 (2.2) |
| | 20-30 | 14820 (61.6) | 3219.4 (485.6) | 441.2 (91.5) | 578 (3.9) | 19.1 (2.1) | 16.5 (1.9) | 21.8 | 59.2 (13.3) | 4.9 (2.3) |
| | >30 | 1321 (5.5) | 3351.1 (531.5) | 467.5 (102.1) | 49 (3.7) | 19.4 (2.2) | 16.7 (2.0) | 21.9 | 61.7 (13.8) | 5.1 (2.3) |
| | Unknown | 1853 (7.7) | 3236.4 (513.7) | 442.8 (94.3) | 77 (4.2) | 19.2 (2.1) | 16.2 (1.8) | 23.0 | 59.7 (13.5) | 4.7 (2.3) |
| **Infant gender** | Male | 12298 (51.1) | 3251.3 (496.7) | 439.2 (91.4) | 462 (3.8) | 19.1 (2.1) | 16.5 (1.9) | 21.8 | 60.1 (13.5) | 4.9 (2.3) |
| | Female | 11763 (48.9) | 3133.9 (481.5) | 434.3 (92.1) | 468 (4.0) | 19.0 (2.1) | 16.4 (1.9) | 21.8 | 57.6 (12.9) | 4.8 (2.3) |

**Table 2. Descriptives of the placental measures (N=24,061).**

| | Overall Population | |
|---|---|---|
| | **Mean (SD)** | **Range** |
| **α (exponentiated)** | 1.03 (1.18) | 0.38, 2.42 |
| **β** | 0.78 (0.02) | 0.66, 0.89 |
| **Umbilical cord length** | 58.8 (13.4) | 7,98 |
| **Cord insertion to margin** (cm) | 4.8 (2.3) | -13, 13 |
| **Largest diameter of placenta** (cm) | 19.0 (2.1) | 12, 30 |
| **Smallest diameter of placenta** (cm) | 16.5 (1.9) | 9, 25 |
| **Chorionic plate area** ($cm^2$) | 247.7 (50) | 77,569 |
| **Placental thickness** (cm) | 2.2 (0.5) | 0.4, 4.5 |
| **Placental weight** (g) | 437 (92) | 90, 1100 |
| **Birthweight** (g) | 3194 (493) | 1219, 5613 |

**Table 3. Placental, maternal and fetal influences on α**

| **Variable** | | **Multivariate model- Placental variables only (N=24,061)** | **Multivariate model- Maternal and fetal variables only (N=21,603)** | **Multivariate model – All variables (N=21,603)** |
|---|---|---|---|---|
| | | | | |
| **Placental shape** | **Round-oval (23,131)** | **-0.021 (0.005)***** | | **-0.020 (0.005)***** |
| | **Other than round/oval (930)** | | | |
| **Chorionic plate area** | | **-0.001 (0.000)*** | | ***-0.001 (0.000)****** |
| **Disk ellipsivity** | | ***0.17 (0.004)****** | | ***0.16 (0.03)****** |
| **Larger diameter** | | ***0.017 (0.004)****** | | ***0.030 (0.004)****** |
| **Smaller diameter** | | ***0.042 (0.004)****** | | ***0.054 (0.004)****** |
| **Disc thickness** | | ***0.010 (0.000)****** | | ***0.012 (0.000)****** |
| **Cord length** | | ***0.001 (0.000)****** | | ***0.001 (0.000)****** |
| **Relative cord eccentricity** | | **0.014 (0.007)*** | | 0.008 (0.007) |
| **Maternal age** | | | 0.000 (0.000) | **-0.001 (0.000)**** |
| **Parity** | | | 0.000 (0.000) | **0.001 (0.001)*** |
| **Smoking** | | | ***0.022 (0.001)**** | ***0.017 (0.001)**** |
| **Infant gender** | | | ***0.020 (0.002)**** | ***0.017 (0.002)**** |
| **Birth length** | | | ***0.003 (0.000)**** | ***-0.011 (0.000)**** |
| **Maternal BMI** | | | ***-0.001 (0.000)**** | ***-0.001 (0.000) **** |
| **Socioeconomic status** | | | 0.000 (0.001) | ***-0.007 (0.001)**** |
| **African-American race** | | | 0.002 (0.003) | ***0.022 (0.002)**** |
| **Gestational age** | | | ***-0.007 (0.001)**** | ***-0.009 (0.001)**** |

***** P<0.0001 bolded and italicized; **P<0.001; *P<0.05; Not bolded, P>0.05.**

**Table 4. FPR and scaled FPR (using ¾ power) .**

| FPR (SD)<br>FPR scaled (SD) | Chorionic Plate Area | | |
|---|---|---|---|
| | **Lowest Quartile** | **Mid Quartiles** | **Upper Quartile** |
| **Thickness** | | | |
| **<2.0 cm** | **8.46 (1.50)** | **7.84 (1.23)** | **7.36 (1.23)** |
| | **1.14 (.26)** | **1.04 (0.19)** | **0.95 (0.19)** |
| | | | |
| **2.0-2.5 cm** | **8.03 (1.31)** | **7.49 (1.11)** | **6.95 (1.03)** |
| | **1.07 (0.21)** | **0.98 (0.17)** | **0.89 (0.16)** |
| | | | |
| **>2.5 cm** | **7.34 (1.20)** | **6.83 (1.00)** | **6.33 (1.00)** |
| | **0.97 (0.21)** | **0.88 (0.16)** | **0.80 (0.15)** |